\documentclass[10pt]{article}
\usepackage[english]{babel}
\usepackage[latin1]{inputenc}
\usepackage{amsfonts,amssymb}
\usepackage{amsmath}
\usepackage{cite}
\usepackage{graphicx}
\usepackage{color}

\begin{document}

\title{EFFECTIVE FREE ENERGY FOR INDIVIDUAL DYNAMICS}

\author{\footnotesize SEBASTIAN GRAUWIN, DOMINIC HUNT, ERIC BERTIN, PABLO JENSEN\\
\footnotesize Universit\'e de Lyon, Laboratoire de Physique, ENS Lyon and CNRS,\\ 
\footnotesize 46 all\'ee d'Italie, F-69007 Lyon, France\\
\footnotesize IXXI Complex Systems Institute, 5 rue du Vercors, F-69007 Lyon, France\\
\footnotesize corresponding author: pablo.jensen@ens-lyon.fr}

\date{}

\maketitle


\begin{abstract}
Physics and economics are two disciplines that share the common challenge of linking microscopic and macroscopic behaviors. However, while physics is based on collective dynamics, economics is based on individual choices. This conceptual difference is one of the main obstacles one has to overcome in order to characterize analytically economic models. In this paper, we build both on statistical mechanics and the game theory notion of Potential Function to introduce a rigorous generalization of the physicist's free energy, which includes individual dynamics. Our approach paves the way to analytical treatments of a wide range of socio-economic models and might bring new insights into them. As first examples, we derive solutions for a congestion model and a residential segregation model.
\end{abstract}

\section{Introduction}
The intricate relations between the individual and collective levels are at the heart of many natural and social sciences \cite{grauwin2009}. Physics looks at the collective level, selecting the configurations that minimize the global free energy \cite{goodstein}. 
In contrast, economic agents behave in a selfish way, and equilibrium is attained when no agent can increase its own satisfaction \cite{mascolell}. 

Recently, physicists have tried to use statistical physics approaches to understand social phenomena such as the segregation 
transition \cite{dallasta,vinkovic06}. The idea seems promising because statistical physics has successfully bridged the gap between the micro and macroscopic levels for physical systems governed by collective dynamics. However, progress remained slow due to the lack of an appropriate framework allowing to take into account the selfish dynamics typical of socio-economic agents. On the other hand, game theorists have developed in the last decades the notion of Potential Games, in which each player's gain resulting from a change of state is equal to the variation of a potential function \cite{anderson92,young93,monderer96,ui2000svr}. This potential function hence provides what physicists need: a link between individual and collective levels. 

In this paper, we propose a generic analytical framework that builds on concepts originating both from statistical physics and game theory. We introduce a rigorous generalization of the physicist's free energy, which encompasses individual dynamics. By introducing a ``link'' state function that is maximized in the stationary state, we pave the way to analytical treatments of a much wider class of systems, where dynamics is governed by individual strategies. Quantitative solutions of two models are also provided as examples.


\section{The Model}

\subsection{Generic model}

Our model represents in a schematic way the dynamics of agents making individual choices.
Throughout this paper, $\mathcal{N} = \{1,\ldots,N\} $ denotes a finite set of agents, $q_i \in \{1,\ldots,Q\}$ the choice 
of agent $i \in \mathcal{N}$, the vector $\vec{q} = (q_i)_{i\in\mathcal{N}}$ describing the state of the system 
and $\mathcal{N}_q$ (resp. $n_q$) the set (resp. number) of agents following choice $q \in \{1,\ldots,Q\}$.
Each agent $i \in \mathcal{N}$ can moreover be characterized by his utility function, which describes the degree of satisfaction concerning his choice. 
An agent's utility function is supposed to depend only on his own choice and
on the set $\mathcal{N}_{q_i}$ of agents making the same choice, namely 
$u_i(\vec{q})=u_i(q_i,\,\mathcal{N}_{q_i})$.
We also introduce the collective utility, defined as the total utility of all the agents: $U(\vec{q})=\sum_i\,u_i(\vec{q})$.

The dynamical rule allowing the agents to change their choice is the following. At each time step, one picks up at random an agent and a choice $q^*\in\{1,\ldots,Q\}$. Then the agent goes from choice $q_i$ to choice $q_i'=q^*$ with probability:  
\begin{equation}
P_{q_i \rightarrow q_i'} = \frac{1}{1+e^{-\Delta u_i/T}},
\end{equation}
where $\Delta u_i=u_i(\vec{q'}) - u_i(\vec{q})$ is the variation of the agent's own utility upon his change of choice.
The parameter $T>0$ is a ``temperature'' that introduces in a standard way some noise on the decision process \cite{anderson92}. It can be interpreted as the effect of features that are not explicitly included in the utility function but still affect the decision.

We wish to find the stationary probability distribution $\Pi(\vec{q})$ of the microscopic configurations $\vec{q}$. 
If $\Delta u_i$ can be written as $\Delta u_i=\Delta L \equiv L(\vec{q'})-L(\vec{q})$, where $L(\vec{q})$ is a state function of the configuration $\vec{q}$, then the dynamics satisfies detailed balance \cite{evans} and the distribution 
$\Pi(\vec{q})$ is given by
\begin{equation}
\Pi(\vec{q}) = \frac{1}{Z}\, e^{L(\vec{q})/T},
\label{eq2}
\end{equation}
with $Z$ a normalization constant.

It can be shown \cite{grauwin2009,ui2000svr} that a sufficient and necessary condition\footnote{If the state function $L$ exists, the relation $\Delta u_i=\Delta L$ requires only that it be defined up to a constant.} for a ``linking'' function $L$ to exist is 
\begin{equation}
u_i(q_i,\,\mathcal{N}_{q_i}\setminus \{j\})-u_i(q_i,\,\mathcal{N}_{q_i}) = 
u_j(q_j,\,\mathcal{N}_{q_j}\setminus \{i\})-u_j(q_j,\,\mathcal{N}_{q_j})
\label{cond}
\end{equation}
for any $i \in \mathcal{N}$ and $j \in \mathcal{N}$. Note that this relation is automatically satisfied in the case $q_i \neq q_j$. Eq.~(\ref{cond}) expresses a symmetric condition on the utility variation (or {\it externality}) an agent produces on another one when he changes his choice. Condition Eq.~(\ref{cond}) is also rather easy to satisfy in case of homogeneous agents sharing the same utility function (see examples in section \ref{appl}). In contrast, this condition imposes more restriction to models with heterogeneous agents and explicit examples are then more difficult to build.

Interestingly, the linking function $L$ appears as (the opposite of) an effective energy in terms of physical systems (Eq.~(\ref{eq2}) being the analogue of a Gibbs distribution), but also corresponds to the notion of potential function in game theory \cite{monderer96,ui2000svr}.


\subsection{Homogeneous agents}
In the following, we restrict our study to models where the agents' utility functions can be written as $u_i(\vec{q})=u(q_i,n_{q_i}/H)$, where $H$ is a parameter characterizing the typical number of agents making a given choice (for instance the natural capacity of an infrastructure). This parameter is assumed to scale linearly with $N$, the ratio $h=H/N$ being fixed. This particular form of utility function implies that the agents share homogeneous properties and that they are sensitive to the relative proportion of agents making the same choice as them. It also implies that Eq (\ref{cond}) is verified, meaning that a linking function $L$ always exists. It is straightforward to check that it can be written as :
\begin{equation}
L(\vec{q})=\sum_{q =1}^{Q} \sum_{m=0}^{n_q(\vec{q})} u(q,m/H)
\end{equation}
In the limit $N \rightarrow \infty$ with $\rho_q=n_q/H$ fixed, one finds
\begin{equation}
L(\vec{q})/N \quad \rightarrow \quad h \sum_{q =1}^{Q}\int_{0}^{\rho_q}\!\! u(q,\rho)\,d\rho.
\end{equation}
This particular form of the potential function $L$ allows us to interpret it as the sum of the individual marginal utilities gained by agents as they progressively enter the system after leaving a reservoir of zero utility (a similar interpretation can be found in \cite{mascolell2}). 

Since the agents are supposed to be identical (but still distinguishable), it seems natural
to keep track of ``mesoscopic'' observables such as the coarse-grained states $x \equiv \{\rho_q\}$ rather than the ``microscopic'' states $\vec{q}$. The number of states $\vec{q}$ corresponding to a given coarse-grained state $x$ is quantified by its logarithm $S(x)$, where:
\begin{equation}
S(x)/N =  \frac{1}{N}\ln{\frac{N!}{\prod_{q} n_q!}} \quad \xrightarrow[N \rightarrow \infty]{h,\,\{\rho_q\} \mbox{ fixed}} \quad 
-\ln{h} - h\sum_{q =1}^{Q} \rho_q \ln\rho_q.
\end{equation}
The stationary distribution of the coarse-grained configurations hence takes the form:
\begin{equation}
\Pi_{N,T}(x) = \frac{1}{Z_{N,T}}\, e^{F(x)/T}=\frac{1}{Z^{'}_{N,T}}\, e^{(hN/T)\sum_q f_{N,T}(q,\rho_q)}
\label{piinf}
\end{equation}
where $F(x)\equiv L(x)+TS(x)$ can be seen as (the opposite of) an effective free energy of the system, $Z^{'}_{N,T}=Z_{N,T}h^{-N}$ and where
\begin{equation}
f_{N,T}(q,\rho) = \sum_{m=0}^{\rho H} u(q,m/H) - T\rho\ln{\rho}
\end{equation}
In the limit $N \rightarrow \infty$ with $\rho_q=n_q/H$ fixed, one finds
\begin{equation}
f_{N,T}(q,\rho)  \quad \rightarrow \quad f_{\infty,T}(q,\rho) 
\equiv \int_{0}^{\rho}\!\! u(q,\rho')\,d\rho' - T\rho \ln\rho
\end{equation}

According to the form of the distribution $\Pi_{N,T}$ given by Eq (\ref{piinf}), in the limit of large $N$ the stationary configurations are those that maximize the sum $\sum_q f_{\infty,T}(q,\rho_q)$ under the constraint $h\sum_q \rho_q = 1$. We 
explore different maximization procedures in examples of applications presented in next section.


\section{Applications}
\label{appl}
\subsection{Road congestion}

We apply here our generic model framework to a simplified version of Chu's \cite{chu1995} congestion model. In this model, a number $N$ of identical commuters travel every morning from home to work. All agents travel on the same road and wish to arrive at time $t^*$. Since congestion is a collective phenomenon that no single agent can master to arrive at her preferred arrival time $t^*$, agents have to choose a less optimal arrival time $t \in \mathbb{Z}$ (time is supposed to be discrete) in order to minimize the private trip cost, $c(t)$, which includes two parts. The first part is the travel time cost $\alpha TT(t)$ where $\alpha$ is the unit cost of travel time and $TT(t)$ is the travel time. The second part is the schedule delay cost, which is $\beta (t^*-t)$ if one arrives early and $\nu (t-t^*)$ if one arrives late, $\beta$ and $\nu$ being unit costs of schedule delay. To make analytical calculations possible, the travel time is supposed to depend on the number $n_t$ of commuters arriving at time $t$ through the function $TT(t) = (n_t/H)^{\gamma}$ where $H$ and $\gamma$ are fixed parameters. The parameter $H$ can be interpreted as a standard road capacity (the linearity between $H$ and $N$ can thus reflect that bigger roads are built when the traffic is more important) and the parameter $\gamma$ measures the elasticity of travel time with respect to $n_t$.

\begin{figure}[th]
\begin{center}
\includegraphics[width=10cm]{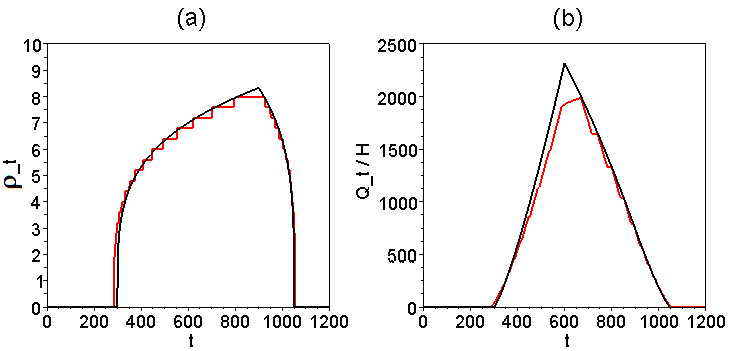}
\end{center}
\vspace*{8pt}
\caption{\textbf{Stationary values} derived by a maximization of $L$ for $N=5000$ (in red) and $N\rightarrow \infty$ (in black) commuters. 
\textbf{(a) Normalized proportion of arriving agents} $\rho_t = n_t / H$.
\textbf{(b) Normalized proportion of agents on the road} $Q_t / H$, where $Q_t$ is the number of agent queueing on the road at a given time $t$. It is computed as the difference between the number of departed and arrived agents $Q_t=\sum_{t'=0}^t \big( d_{t'} - n_{t'}\big)$, where the distribution of departure time $\{d_t \}$ is deduced from the distribution of arrival time $\{n_t\}$ and travel time $\{TT(t)\}$. See \protect\cite{chu1995} for more details on this procedure.
The computations have been realized in the limit $T\rightarrow 0$, with the parameters values $t^*=900$, $\alpha=2$, $\beta=1$, $\nu=4$, $\gamma=4$, $h=4.\, 10^{-4}$.}
\label{congestion}
\end{figure}

This congestion model fits our framework model, the utility of an agent arriving at time $t$ being
\begin{equation}
u(n_t) = - c(t) = - \left\{\begin{array}{ll}
\alpha (n_t/H)^\gamma + \beta |t^*-t| & \mbox{ if } t < t^*\\
\alpha (n_t/H)^\gamma + \nu |t^*-t| & \mbox{ if } t \geq t^*
\end{array}\right.
\end{equation}
The stationary coarse-grained configurations in the limit of large $N$ are thus those that maximize the sum $\sum_t f_{\infty,T}(t,\rho_t)$, where:
\begin{equation}
f_{\infty,T}(t,\rho)  =  \left\{\begin{array}{ll}
\frac{\alpha}{\gamma + 1}\rho^{\gamma +1} + \beta\rho|t^*-t| -T\rho\ln\rho & \mbox{ if } t < t^*\\
\frac{\alpha}{\gamma + 1}\rho^{\gamma +1} + \nu\rho|t^*-t| -T\rho\ln\rho & \mbox{ if } t \geq t^*
\end{array}\right.
\label{fcong}
\end{equation}
under the constraint $h\sum_t \rho_t = 1$. Since the sum  $\sum_t f_{\infty,T}(t,\rho_t)$ is maximized by the stationary configuration, the stationary values of $\{\rho_t\}$ verify (for any $t$ such that $\rho_t \neq 0$) the relation: 
\begin{equation}
\frac{\partial f_{\infty,T}}{\partial \rho}(t,\rho_t) = \frac{\partial f_{\infty,T}}{\partial \rho}(t^*,\rho_{t^*})
\label{cong3}
\end{equation}
which can be easily derived using Lagrange multipliers. Eq~(\ref{cong3}) provides for each time step $t$ an implicit relation between $\rho_t$ and $\rho_{t^*}$ (resp. the normalized densities at time $t$ and $t^*$). All these implicit relations along with the conservation of the number of agents expressed by the condition $h\sum_t \rho_t = 1$ allow one to compute numerically the distribution $\{\rho_t\}$. Fig.~\ref{congestion} presents results that have been computed with this method, for $N \to \infty$ and for a finite value ($N=5000$) of the number of agents obtained by solving numerically an equivalent of Eq.~(\ref{cong3}). These results suggest that finite size effects remain small as soon as $N$ reaches values of a few thousands, and that the main properties of the model can be studied in the limit $N \to \infty$. 
Notice that our analysis of the congestion model in terms of a potential games allows us to determine (numerically) the stationary state of the systems even for finite size, which is rarely possible with usual economics analysis since, quoting  \cite{otsubo2008}, {\it the common practice in transportation science and economics is to use continuous models for analyzing phenomena that are essentially discrete}.


\subsection{Residential choice}

\begin{figure}[th]
\begin{center}
\includegraphics[width=10cm]{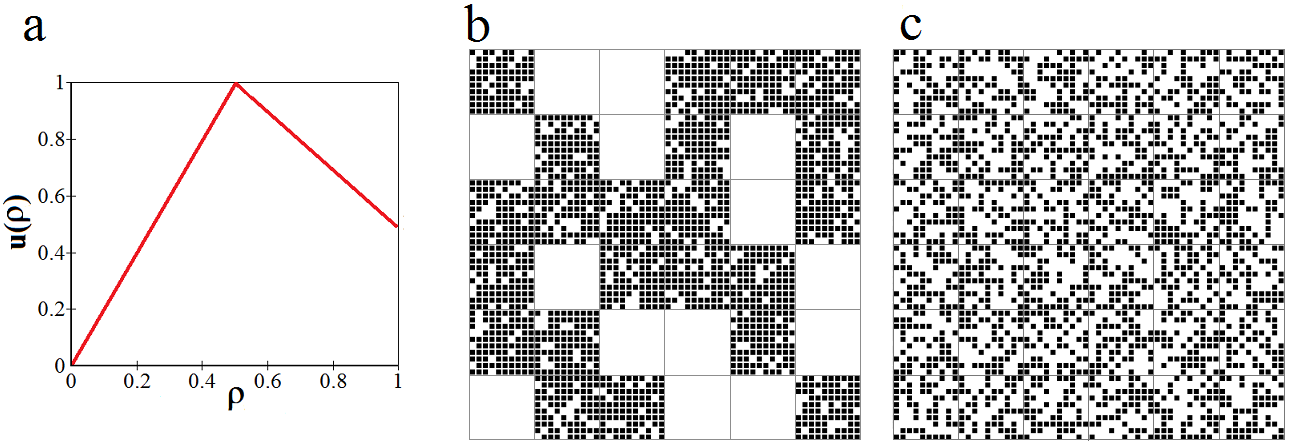}
\end{center}
\vspace*{8pt}
\caption{\textbf{(a)}. Asymmetrically peaked individual utility as a function of block density. The utility is defined as $u(\rho)=2\rho$ if $\rho \leq 1/2$ and $u(\rho)=3/2-\rho$ if $\rho > 1/2$. Agents strictly prefer half-filled neighborhoods ($\rho=1/2$). They also prefer overcrowded ($\rho=1$) neighborhoods to empty ones ($\rho=0$).
\textbf{(b)} Stationary configurations of a virtual city for $T \rightarrow 0$. Blocks are separated in two phases with different densities of agents.
\textbf{(c)} Stationary configurations of a virtual city for $T \gg 1$. Agents are distributed homogeneously between the blocks. 
The city is composed of $Q=36$ blocks containing each $H=100$ cells, with a mean density
$\langle \rho_q \rangle=N/HQ=1/2$.
}
\label{segr1}
\end{figure}

As a second example, we apply our generic framework to a Schelling-like model describing the dynamics of residential moves in a city \cite{grauwin2009,schelling78}. The virtual city is divided into $Q$ blocks ($Q\gg 1$), each block containing $H$ cells or flats (Fig \ref{segr1}). We assume that each cell can contain at most one agent, so that the number $n_q$ of agents in a given block satisfies $n_q \le H$. An agent's utility depends only on the density $\rho_q=n_q/H$ of the block he is living in, ie $u_i(\vec{q})=u(\rho_q)$, with the convention that $u(\rho)=-\infty$ for $\rho > 1$.

Once again, this residential model can be solved thanks to our framework. With the particular choice of the asymmetrically peaked utility function given Fig \ref{segr1}(a), the stationary coarse-grained configurations in the limit of large $N$ are thus those which maximize the sum $\sum_q f_{\infty,T}(q,\rho_q)$, where\footnote{Caution: the expression of the entropy $S(x)$ depends on the precise definition of the ``microscopic'' states of the system. If the agents were to choose the \emph{cell} (instead of the \emph{block}) in which they would like to move, one would need to add the term $S'(x)=\sum_q\ln\frac{H!}{(H-n_q)!}$ in the expression of the entropy.} :

\begin{equation}
f_{\infty,T}(q,\rho)  =  \left\{\begin{array}{cl}
\rho^2 -T\rho\ln\rho & \mbox{ if } \rho < 1/2 \\
-\rho^2/2 + 3 \rho/2-3/8 -T\rho\ln\rho & \mbox{ if } \rho \geq 1/2
\end{array}\right.
\label{fsegr}
\end{equation}

To perform the maximization procedure, one can follow standard physics methods used in the study of phase transitions (like liquid-vapor coexistence \cite{callen85}). In this case, it is a simple exercise to determine that for $N \gg 1$ and in the limit $T\rightarrow 0$, a phase separation occurs. A fraction $\min\big(2\sqrt{3}/3hQ,\,1\big)$ of the blocks have a density $\min\big(\sqrt{3}/2,\,1/hQ\big)$ while the others are empty. This result is illustrated on Fig.~\ref{segr1}. For more details on the maximization procedure, the interested reader is referred to \cite{grauwin2009}.


\section{Discussion}
We derived in this paper an effective free energy $F=L+TS$ in a generic framework model. The main property of this function is to intimately connect the individual and global points of view. Our simple model raises a number of interesting questions: 
in the limit $T \to 0$, the stationary configurations are those maximizing the potential $L$ and not the collective utility $U$. Hence, they may differ from the simple collection of individual optima \cite{kirman92}, illustrating the unexpected links between micromotives and macrobehavior.   

More specifically, we derived a simple expression of this effective free energy in the thermodynamic limit $N\to \infty$ for homogeneous agents. We showed that this simple form allows us to derive in an easy and flexible way quantitative solutions to economics models based on individual dynamics. This approach can be extended to some models with heterogeneous agents. Possible examples range from a two-population residential segregation model \cite{grauwin2009}, Ising-like model with heterogeneous pairs interactions \cite{nadal2005physique}, or the Hopfield model \cite{ui2000svr}.


\begin{thebibliography}{20}

\bibitem{grauwin2009}
Grauwin S, Bertin E, Lemoy R, Jensen P (2009)
Competition between collective and individual dynamics,
{\it Proc Natl Acad Sci USA} 106: 20622--20626.

\bibitem{goodstein}
Goodstein D (1985) {\it {States of Matter}} (Dover Publications).

\bibitem{mascolell}
Mas-Colell A, Whinston MD, Green JR (1995) {\it Microeconomic Theory} (Oxford University Press, NY).

\bibitem{dallasta}
Dall'Asta L, Castellano C, Marsili M (2008)
Statistical physics of the Schelling model of segregation
{\it J Stat Mech}, L07002

\bibitem{vinkovic06}
Vinkovic D, Kirman A (2006)
A physical analogue of the Schelling model.
{\it Proc Natl Acad Sci USA} 103:19261--19265.

\bibitem{anderson92}
Anderson SP, De~Palma A, Thisse JF (1992) {\it {Discrete Choice Theory of Product Differentiation}} (MIT Press).

\bibitem{young93}
Young HP (1993) 
The evolution of conventions,
{\it Econometrica} 61:57--84.

\bibitem{monderer96}
Monderer D, Shapley L (1996) 
Potential Games,
{\it Game Econ Behavior} 14:124--143.

\bibitem{ui2000svr}
Ui T, (2000)
A Shapley value representation of potential games,
{\it Games Econ Behavior} 31:125--135.

\bibitem{evans}
Evans M, Hanney T (2005) 
Nonequilibrium statistical mechanics of the zero-range process and related models,
{\it J Phys Math Gen} 38:R195--R240.

\bibitem{mascolell2}
Hart S, Mas-Colell A (1989)
Potential, Value, and Consistency,
{\it Econometrica} 57(3):589--614.

\bibitem{callen85}
Callen H (1985) {\it Thermodynamics and an introduction to thermostatistics}, eds Wiley J and sons (New York).

\bibitem{chu1995}
Chu X (1995)
Endogenous trip scheduling: the Henderson approach reformulated and compared with the Vickrey approach,
{\it J. Urban Econ.} 37: 324--343.

\bibitem{otsubo2008}
Otsubo H, Rapoport A (2008)
Vickrey's model of traffic congestion discretized,
{\it Transportation Research Part B} 42: 873--889.

\bibitem{schelling78}
Schelling TC (1978) {\it Micromotives and macrobehavior} (Norton, New York).

\bibitem{kirman92}
Kirman AP (1992) 
Whom or What Does the Representative Individual Represent?
{\it J Econ Perspect} 6:117--36.

\bibitem{nadal2005physique}
Nadal J.P., Gordon M.B. (2005)
Physique statistique de ph\'enom\`enes collectifs en sciences \'economiques et sociales,
{\it Math\'ematiques et sciences humaines} 172:67--89.

\end{thebibliography}
\end{document}